\documentclass[journal]{IEEEtran}

\usepackage{amsthm,amssymb,amsmath,mathtools}
\usepackage{subfig}
\usepackage{multirow}
\usepackage{graphicx,caption}
\usepackage{float}
\usepackage[ruled,vlined,linesnumbered]{algorithm2e}
\usepackage{xr}
\makeatletter

\begin{document}

\makeatletter \global\long\def\specificthanks#1{\@fnsymbol{#1}}
\makeatother

\title{Cancellable Template Design for Privacy-Preserving EEG Biometric Authentication Systems}

\author{Min Wang,~\IEEEmembership{Member,~IEEE,}
        Song Wang,
        Jiankun Hu,~\IEEEmembership{Senior Member,~IEEE}

\thanks{Min Wang and Jiankun Hu are with the School of Engineering and Information Technology, University of New South Wales, Australia (e-mail: maggie.wang1@adfa.edu.au; j.hu@adfa.edu.au). Song Wang is with the School of Computing, Engineering and Mathematical Sciences, La Trobe University, Australia (e-mail: song.wang@latrobe.edu.au)}
}

\maketitle

\begin{abstract}
As a promising candidate to complement traditional biometric modalities, brain biometrics using electroencephalography (EEG) data has received a widespread attention in recent years. However, compared with existing biometrics such as fingerprints and face recognition, research on EEG biometrics is still in its infant stage. Most of the studies focus on either designing signal elicitation protocols from the perspective of neuroscience or developing feature extraction and classification algorithms from the viewpoint of machine learning. These studies have laid the ground for the feasibility of using EEG as a biometric authentication modality, but they have also raised security and privacy concerns as EEG data contains sensitive information. Existing research has used hash functions and cryptographic schemes to protect EEG data, but they do not provide functions for revoking compromised templates as in cancellable template design.
This paper proposes the first cancellable EEG template design for privacy-preserving EEG-based authentication systems, which can protect raw EEG signals containing sensitive privacy information (e.g., identity, health and cognitive status). A novel cancellable EEG template is developed based on EEG graph features and a non-invertible transform. The proposed transformation provides cancellable templates, while taking advantage of EEG elicitation protocol fusion to enhance biometric performance. The proposed authentication system offers equivalent authentication performance (8.58\% EER on a public database) as in the non-transformed domain, while protecting raw EEG data. Furthermore, we analyze the system's capacity for resisting multiple attacks, and discuss some overlooked but critical issues and possible pitfalls involving hill-climbing attacks, second attacks, and classification-based authentication systems.
\end{abstract}


\IEEEpeerreviewmaketitle{}

\section{Introduction}
\begin{figure}[h]
\centering
\includegraphics[width=0.8\columnwidth]{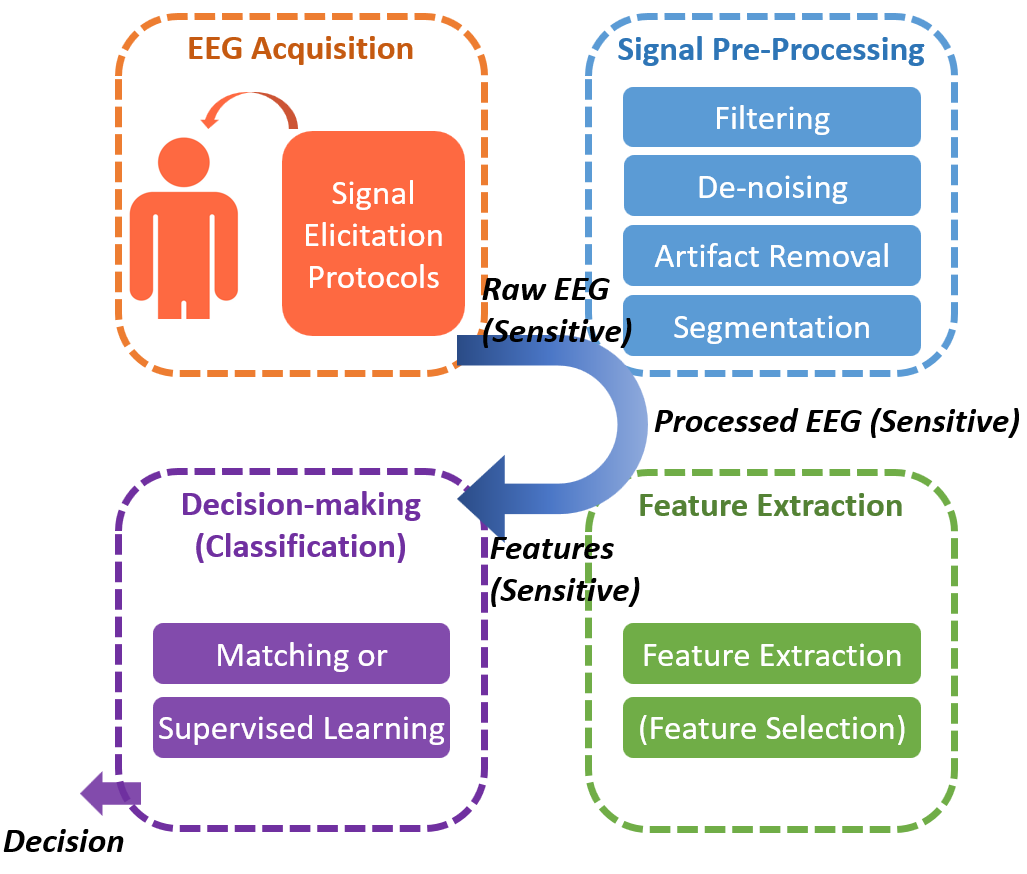}
\caption{A typical structure of EEG-based biometric systems.}
\label{fig:eeg_system}
\end{figure}

Conventional biometric techniques such as fingerprint and face recognition share vulnerabilities in terms of confidentiality and robustness against circumvention~\cite{hadid2015biometrics} since these biometric traits are observable and can be illegally obtained or forged without the user's awareness, e.g., via high resolution photography~\cite{marasco2014survey}. The need for stronger security has given birth to brain biometrics based on electroencephalography (EEG) signals. Meanwhile, the rapid development of brain-computer interface, neuroscience, and sensor technology has created an environment where EEG is readily available for biometric applications. Potential advantages of EEG biometrics include its robustness against circumvention, support for liveness detection, continuous authentication, and cognitive information indicators~\cite{gui2019survey,wang2019convolutional}. A typical EEG-based biometric recognition system consists of four major modules: signal acquisition, pre-processing, feature extraction, and decision-making, as illustrated in Fig.~\ref{fig:eeg_system}. During data acquisition, EEG signals, captured by sensors from the user's scalp while he or she engages with the elicitation protocol, are transmitted to the processing unit. Since raw EEG data are usually contaminated with noise and artifacts, it is necessary to preprocess the raw data to enhance signal quality. Then discriminant features are extracted from the preprocessed EEG and fed into a decision-making module.

The acquisition of EEG biometrics requires to specify the corresponding signal elicitation protocols, among which the resting protocol is favorable due to its convenience and minimum requirements for data collection. This protocol is widely used as a baseline for calibrating user EEG in brain-computer interface (BCI) systems~\cite{cohen2014analyzing}. More importantly, ongoing EEG under the resting state protocol does not involve external stimulation to or active response from the user during data acquisition, thus minimizing the impact of cognitive state changes on signal stability~\cite{la2012eeg,la2014human}. It also supports operation in a continuous and unobtrusive manner. Alternative protocols include the volitional tasks, e.g., the pass-thoughts and various event-related potential (ERP) protocols. A representative ERP protocol for EEG biometrics is the CEREBRE system~\cite{ruiz2016cerebre}, which uses a series of visual stimuli to elicit individually unique responses from users' multiple functional brain systems. This type of research focuses on how to control users' cognitive states with repetitive sensory stimulation and cognitive tasks to elicit desired distinctive brain responses. The preprocessing, feature extraction, and decision-making modules involve corresponding computational methods. Specifically, the decision-making is achieved by template matching or classification based on supervised learning models. Template matching is adopted in many studies as an effective and solid decision-maker in both identification and authentication scenarios~\cite{la2014human,ruiz2016cerebre,maiorana2015permanence,wang2020brainprint}. Other studies view identification or authentication as a classification problem and explore different machine learning models, such as discriminant analysis~\cite{riera2008unobtrusive,yang2018task}, support vector machines, and neural networks~\cite{wang2019convolutional,debie2021session}. 

However, user authentication is not merely a classification problem, but also entails security considerations. Classification accuracy does not necessarily reflect real biometric performance. It is not theoretically sound to directly apply the standard evaluation procedure of testing a classification model to assess a classification-based authentication system. Unfortunately, many studies fail to differentiate the two concepts. We will discuss this issue in Section~\ref{sec:security}. In addition, there is a need for a unified evaluation of different features and methods in a rigorous user authentication experimental setup.

The contributions of this study are summarized as follows:
\begin{itemize}
    \item A cancellable template design is proposed to attain a privacy-preserving EEG-based authentication system. 
    \item An innovative transformation is designed to generate cancellable templates based on EEG graph features and a non-invertible transform. The proposed transformation is tailored for EEG biometrics allowing for elicitation protocol fusion to enhance matching performance, while providing template protection.
    \item A new concept of second attacks is introduced to examine the possibility of breaking into a system using pre-obtained solutions after the system has revoked the compromised template. 
    \item Pre-image and hill-climbing attacks are widely used criteria to assess cancellable biometrics. We reveal that these criteria do not fully apply to security assessment of cancellable biometrics. Hence, we re-define the concept of pre-image attacks suitable for cancellable biometrics. 
    \item Extensive experiments are carried out to evaluate the effects of pre-image and hill-climbing attacks. The results demonstrate that cancellable template design based on many-to-one mapping is inherently resistant to these attacks, which is contrary to the common understanding in the field. 
    \item In-depth analysis is conducted on pitfalls involved in the evaluation procedure of supervised learning-based authentication systems.
\end{itemize}

The rest of this paper is organized as follows. Section~\ref{sec:related} reviews the state of the art on EEG biometrics and protection mechanisms. Section~\ref{sec:method} presents the proposed methodology. Section~\ref{sec:experiment} describes the experimental design, followed by results in Section~\ref{sec:result} and security analysis and discussion in Section~\ref{sec:security}. The conclusion and future directions are summarized in Section~\ref{sec:conclusion}.

\section{Related Work}\label{sec:related}

\subsection{EEG Biometrics}
EEG under the resting state has been investigated for biometric applications for over a decade and recent studies on the permanence issue suggested that the resting state protocol presents an effective and robust condition for biometric recognition~\cite{la2012eeg,maiorana2015permanence}. In the resting state elicitation protocol, the user remains relaxed with eyes closed (EC) or eyes open (EO) without performing any particular task. The rationale behind it, besides its implementation simplicity, is the neurophysiological evidence which indicates that ongoing EEG under the resting state carries unique identity information (e.g., those related to heritability and personality factors)~\cite{campisi2014brain}. In addition, EEG signals present large intra-user variations that could hinder the biometric performance. In order to improve system performance and robustness, the fusion of multiple elicitation protocols is adopted. In many works, this is achieved by decision-level fusion through voting schemes~\cite{ruiz2016cerebre}. 
Another way for protocol fusion is to mix the EEG data collected under different elicitation protocols to create a data set that contains the generalized unique pattern of each user~\cite{debie2021session}. 
This strategy has been adopted in many EEG biometrics studies to account for the intra-class variability, especially methods based on supervised learning models~\cite{yang2018task}.

Regarding feature extraction, different methods are proposed considering the various characteristics of EEG signals. Based on whether the relationship information between signals of different channels is captured or not, we can categorize EEG features into univariate features and bivariate features. The univariate features are extracted from single channels of signals considering signal characteristics in the time and frequency domains. Popular ones include the coefficients of autoregressive (AR) models~\cite{maiorana2015permanence,nakamura2017ear}, fuzzy entropy~\cite{mu2016eeg}, and power spectral density (PSD) features~\cite{maiorana2015permanence,la2014human}, which reflect time-dependency, dynamic complexity, and spectral characteristics of EEG, respectively. 

On the other hand, the bivariate features are based on brain connectivity which captures the interactive or structural information between EEG channels. Different statistical and effective metrics have been used for establishing connectivity between EEG channels, including the Pearson's correlation~\cite{wang2019convolutional,fraschini2019robustness}, Granger causality~\cite{min2017individual}, spectral coherence~\cite{la2014human}, and phase synchronization indices~\cite{fraschini2019robustness,wang2020brainprint}. Moreover, graph features extracted from the brain connectivity networks are also proposed for EEG biometrics~\cite{fraschini2014eeg,wang2020brainprint}. Recent findings suggest that, compared with univariate features, bivariate features are more robust against the intra-user variations across sessions, thus improving biometric performance~\cite{wang2019convolutional,wang2020brainprint}. The result also shows that the phase synchronization, especially the $\rho$ index, is a sound metric to estimate EEG connectivity for biometric applications.

For classification in EEG biometrics, existing methods can be categorized into matching-based classification and supervised learning-based classification. In authentication, the matching-based methods predict the class label (genuine user or impostor) of a query sample by calculating its similarity to one or multiple registered samples of the claimed user. The similarity was defined by the Euclidean distance~\cite{fraschini2014eeg}, Mahalanobis distance~\cite{la2014human}, Manhattan distance~\cite{maiorana2015permanence}, cosine similarity~\cite{maiorana2015permanence,nakamura2017ear}, and cross-correlation~\cite{ruiz2016cerebre}. Template matching is straightforward and computationally fast, yielding interpretable results. The performance depends on the discriminative capacity of the template. Recent studies also explored different machine learning algorithms for classification in EEG biometrics. Popular classifiers include the linear discriminant analysis (LDA)~\cite{nakamura2017ear,riera2008unobtrusive,yang2018task}, support vector machines (SVMs)~\cite{nakamura2017ear}, and deep neural networks such as multilayer perceptron (MLP) and convolutional neural networks (CNNs)~\cite{wang2019convolutional,debie2021session}. In these methods, training is an essential step that fits the model to a training dataset. The performance of the model not only depends on the capability of the model itself, but also relies on the training procedure and a good training dataset. 

\subsection{Privacy-preserving Mechanisms}
Non-invertible transformation design for biometric systems renders a vital privacy-preserving mechanism for biometric template protection. This type of method applies a one-way transformation to biometric data such that an adversary cannot retrieve the original biometric data, even if the stored template is compromised. The matching or classification of the enrolled template and the query is carried out in the transformed domain to protect the original biometric data from leakage. He~\textit{et al.}~\cite{he2009hashing} studied the potential of hashing EEG features for authentication. Multi-variate autoregressive coefficients were extracted as features from multi-channel EEG signals and then hashed by the fast Johnson-Lindenstrauss algorithm to obtain compact hash vectors. A naive Bayes probabilistic model was used for decision-making based on the EEG hash vectors. Applying cryptographic hashing to biometrics induces variation, as any slight change to the input would completely alter the hash value produced. Bajwa~\textit{et al.}~\cite{bajwa2016neurokey} proposed a key generation method with EEG biometrics. The PSD features were extracted from EEG signals using the discrete Fourier transform and discrete wavelet transform, followed by a Neurokey generation procedure which involves feature selection, binarization and hashing. The term `cancellable' is used in this study to mean that a user's Neurokey can be changed by using the EEG collected in a different cognitive task, if the user's biometric information is compromised. However, such `cancellable biometrics' cannot protect raw EEG data containing sensitive information. Furthermore, the choice of tasks is limited and different tasks would have vastly different performance~\cite{yang2017usability}.
Dama{\v{s}}evi{\v{c}}ius~\textit{et al.}~\cite{damavsevivcius2018combining} developed a cryptographic authentication scheme for EEG biometrics using fuzzy commitment and the error-correcting Bose-Chaudhuri-Hocquenghem codes. Although this method protects data privacy, it is not equipped with cancellability to revoke compromised templates.
Cancellable biometric templates based on non-invertible transforms offer a solution to EEG data protection as well as template revocability. To the best of our knowledge, there has been no cancellable EEG template design reported. In EEG biometric systems, most of the work is based on classification models, where it is infeasible to integrate cancellability. Once the model is compromised, the input can be estimated by genetic algorithms so that the system is cracked.

\begin{figure*}[h]
\centering
\includegraphics[width=0.95\textwidth]{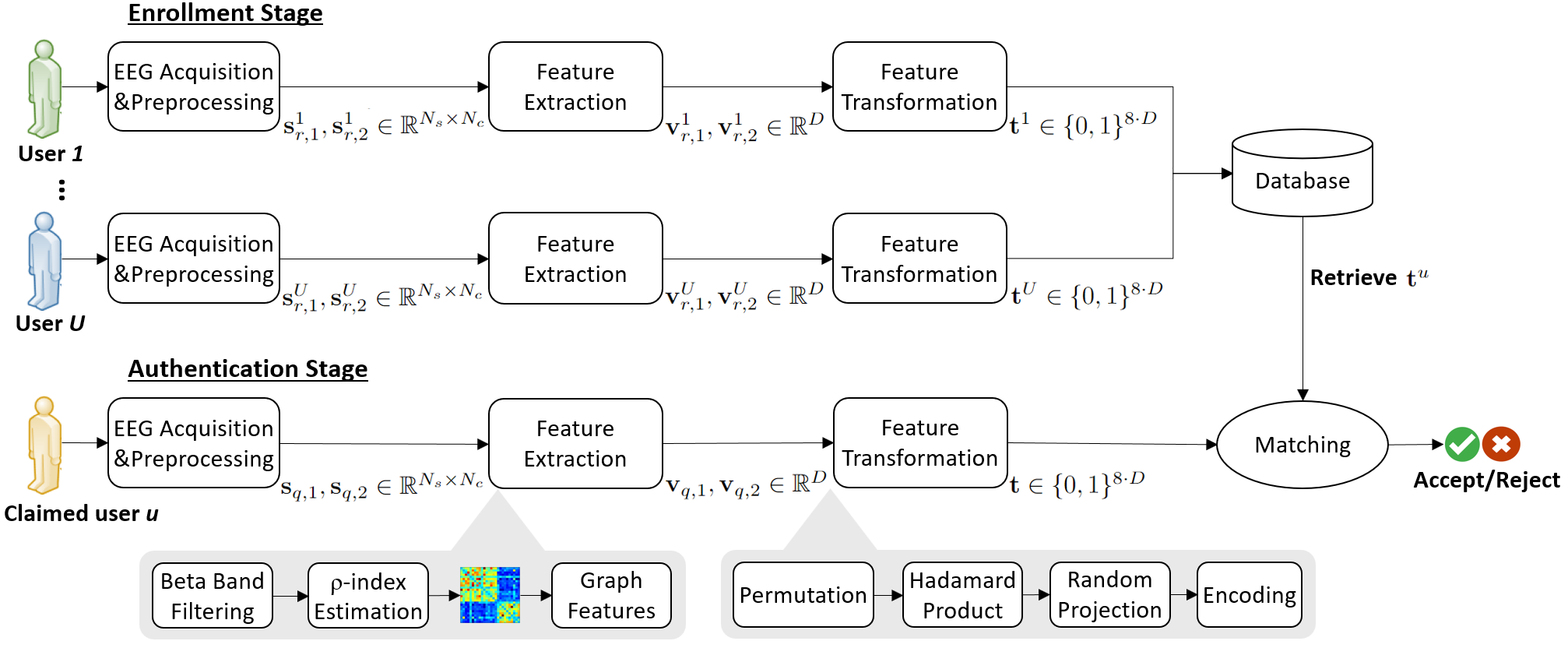}
\caption{The proposed privacy-preserving EEG-based authentication system.}
\label{fig:system}
\end{figure*}

\section{Methodology}\label{sec:method}

In this section, we design a cancellable template to protect EEG biometric data. The proposed privacy-preserving EEG authentication system consists of four main components: signal acquisition, feature extraction, feature transformation, and matching, as illustrated in Fig.~\ref{fig:system}. In the enrollment stage, multi-channel EEG signals are collected from each user under the signal elicitation protocol and fed into the feature extraction module, which constructs brain connectivity graphs and extracts a pair of graph-level features. Then the transformation module takes this pair of features as the input and creates a cancellable template with a set of user-specific parameters. This template is a binary representation, and will be stored in the database. In the authentication stage, a query template is generated following the same procedure and the matching algorithm will output a decision to accept or reject the user. The detailed information of each component is to be presented in the rest of this section, where the innovation mainly lies in the proposed novel transformation, as it exquisitely integrates EEG elicitation protocol fusion with random projection in creating cancellable templates, enhancing both security and matching performance.

\subsection{Signal Elicitation Protocol}\label{sec:acquisition}

\begin{figure}[h]
\includegraphics[width=0.93\columnwidth]{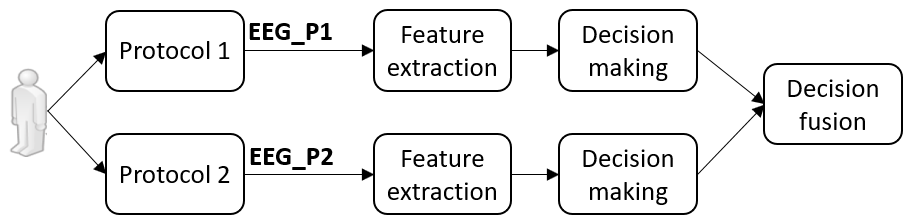}
\includegraphics[width=0.93\columnwidth]{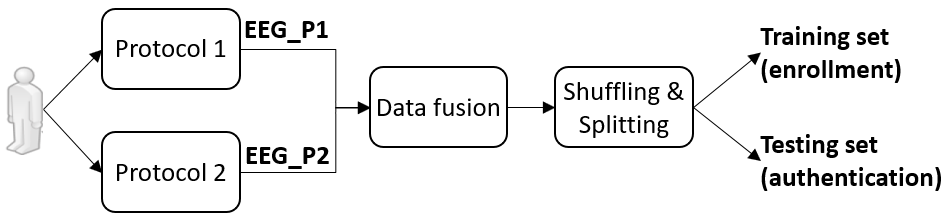}
\includegraphics[width=0.93\columnwidth]{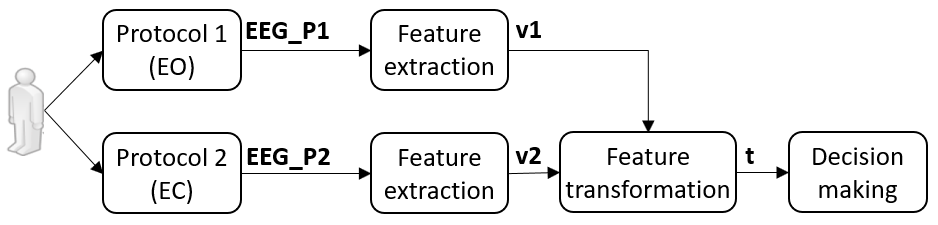}
\caption{Signal elicitation protocol fusion. (1) Elicitation protocol fusion at the decision level; (2) Direct fusion of signals under different elicitation protocols; and (3) The proposed transformation-based elicitation protocol fusion.}
\label{fig:elicitation}
\end{figure}

The resting state EEG elicitation protocol is adopted for signal acquisition. To be specific, two conditions are included, namely the EO and EC states. The user is asked to stay relaxed with eyes closed or eyes open, while the spontaneous EEG signals are recorded. EEG signals present time-varying and non-stationary characteristics, and are sensitive to the cognitive states of the subject, which may affect the biometric performance. Therefore, in order to improve stability, researchers often consider elicitation protocol fusion to get a richer dataset that contains signals in diverse states. We adopt the basic idea of elicitation protocol fusion~\cite{debie2021session}. However, instead of decision-level fusion with majority voting or directly mixing data collected under different protocols to form training and testing sets as in the existing research, as shown is Fig.~\ref{fig:elicitation} (a) and (b), we embed data fusion naturally in the transformation process, as illustrated in Fig.~\ref{fig:elicitation} (c). The benefits of our design are twofold: 1) the entropy of extracted features increases, thus the reliability of the biometric system is enhanced, due to the elicitation protocol fusion; and 2) secure cancellable templates are generated at the same time without extra computational costs. Details of the transformation is presented in Section~\ref{sec:transform}.

\subsection{Feature Extraction}\label{sec:feature}
Feature extraction is based on functional connectivity estimation and graph analysis. For multi-channel EEG signals, the functional connectivity measures the statistical relationship between every two channels. Different statistical metrics can be used, including Pearson's correlation, mutual information, spectral coherence, and phase synchronization measures such as phase locking values. Each connectivity metric reflects different statistical interactions between channels from its unique perspective and defines a peculiar connectivity graph. In our previous study~\cite{wang2020brainprint}, we examined the effectiveness and efficiency of different types of connectivity metrics in constructing EEG connectivity graphs for biometric identification. One important finding is that the $\rho$-index, a phase synchronization measure based on Shannon entropy, is the most effective metric that induces highly distinctive EEG graph patterns among individuals. Moreover, an analysis on the EEG frequency bands revealed that the high frequency range, especially the beta band (13-30 Hz), is the predominant one with individual distinctiveness. Therefore, we perform band-pass filtering on the EEG before functional connectivity estimation to limit the signal to the beta band in a bid to enhance individual uniqueness.

Let $x_i(t)$ and $x_j(t)$ denote two EEG signals (after filtering in the beta band) from channels $i$ and $j$, respectively. The relative phase of $x_i(t)$ and $x_j(t)$ is defined as:
\begin{equation}
    \Delta\phi_r(t) = |\phi_{x_i}(t)-\phi_{x_j}(t)| \mod2\pi ,
\end{equation}
where $\phi_{x_i}(t)$ and $\phi_{x_j}(t)$ are the instantaneous phases of signals $x_i(t)$ and $x_j(t)$, respectively, calculated by Hilbert transform. The $\rho$-index describes the phase synchronization degree of two signals by measuring the deviation of their relative phase from the uniform distribution~\cite{weule1998detection}, expressed as follows:
\begin{align}
	&\rho(x_i,x_j) = \frac{S_{uni}-S}{S_{uni}} \\
	&S=-\sum_{k=1}^N p_k \ln(p_k) ,
\end{align}
where $S$ is the entropy of the relative phase $\Delta\phi_r(t)$ and the probability $p_k$ is approximated by generating a histogram of the relative phase; $S_{uni}$ is the entropy of the uniform distribution (i.e., maximum entropy). The value of $\rho$-index is in the range of $[0,1]$, where $0$ indicates no phase coupling (i.e., uniform distribution of $\Delta\phi_r(t)$) and $1$ indicates perfect phase coupling (i.e., Dirac distributed). For EEG signals of $N$ channels, computing the aforementioned $\rho$-index on every two channels results in an $N\times N$ EEG graph, where each node represents a channel or electrode and the edges indicate the phase synchronization degree of two channels. This graph is undirected with a symmetric adjacency matrix, and reflects the coordinated activation over different brain regions~\cite{sun2012phase}.

Subsequently, nodal and global features are extracted from the $\rho$-index graphs, including node centrality based on the pagerank algorithm and six global graph features; see details below:

\begin{itemize}
\item The pagerank centrality is defined by a random walk of the graph. At each node, the next node is chosen with a probability $p$ from the set of neighbors of the current node. The pagerank centrality score is the average time spent at each node during the random walk.

\item Transitivity is a measure of the degree to which the neighbors of nodes are also connected among themselves. It is defined as $\frac{\sum_{i \in N} 2 t_i}{\sum_{i \in N} k_i(k_i-1)}$. Higher transitivity value means lower segregation in information processing.

\item Modularity quantifies the extent to which a network can be divided into subgroups of nodes with a large number of within-group links and a low number of cross-group links. It is defined as $\frac{1}{l}\sum_{i,j \in N}[w_{ij} - \frac{k_i k_j}{l}] \delta_{m_i,m_j}$, where $l = \sum_{i,j \in N} a_{ij}$ is the number of links, $m_i$ is the module containing node $i$, and $\delta_{m_i,m_j} =1$ if $m_i = m_j$ and $0$ otherwise. A higher modularity value indicates more segregation in information processing.

\item Network characteristic path length $\lambda$ is a measure of routing efficiency between any pair of nodes in the network. It is defined as the average weighted distance between all nodes, i.e., $\lambda = \frac{1}{N(N-1)} \sum_{i \in N} \sum_{i \neq j \in N} min\{\lambda_{ij}\}$, where $ min\{\lambda_{ij}\} $ is the shortest path length between node $i$ and node $j$. A network with high $\lambda$ value has low integration of information.

\item Global efficiency measures the average efficiency of the parallel information transfer in a network, where efficiency is defined by the inverse of the harmonic mean of the shortest path length between two nodes, i.e., $\frac{1}{N(N-1)} \sum_{i \neq j \in N}(\frac{1}{min\{\lambda_{ij}\}})$. Higher global efficiency is equivalent to higher integration of information.

\item Network radius and diameter are the minimal and maximal eccentricity, respectively, where eccentricity of a node is the maximum distance between that node and any other node in the network.
\end{itemize}

Among the six global features, transitivity and modularity measure segregation of the graph; characteristic path length and global efficiency capture the graph integration; and radius and diameter indicate the graph eccentricity. The extracted feature vector has a dimension of $N+6$, where $N$ is the number of channels.

\subsection{Feature Transformation}\label{sec:transform}
Let $\mathbf{v}_1$ and $\mathbf{v}_2$ denote the feature vectors extracted from EEG signals collected under the two elicitation protocols. The proposed transformation takes the input of $\mathbf{v}_1$ and $\mathbf{v}_2$ and generates a secure cancellable template $\mathbf{t}$, as illustrated in Fig.~\ref{fig:elicitation} (c). 

At the enrollment stage, each user is assigned a user key $k$, which is used as the random seed to generate a random permutation of the integers 1 to $N$+6, as follows:
\begin{equation}
    \mathbf{p} = randperm(k,N+6) ,
\end{equation}
where the randperm(seed,integer) is a random permutation function defined on a Pseudorandom number generator (PRNG) which can adopt any generic PRNG algorithm, e.g., the Mersenne Twister algorithm. Then a permuted version of feature vector $\mathbf{v}_1$ is obtained and the Hadamard product of it and $\mathbf{v}_2$ is calculated, as follows:
\begin{align}
\mathbf{v}'_1 &= perm(\mathbf{v}_1,\mathbf{p}) \label{eq:permutation}\\ 
\mathbf{c}    &= \mathbf{v}'_1 \circ \mathbf{v}_2 \label{eq:product}. 
\end{align}

A vector $\mathbf{r}$ is then generated by projecting vector $\mathbf{c}$ with a matrix $\mathbf{M}$, as follows:
\begin{equation}\label{eq:random_projection}
    \mathbf{r} = \mathbf{c} \cdot \mathbf{M} ,
\end{equation}
where $\mathbf{M}$ is a user-specific random projection matrix with more rows than columns to form an underdetermined system of equations, thus making the transformation non-invertible. Finally, the real-valued vector $\mathbf{r}$ is encoded into a binary template $\mathbf{t}$ through the 8-bit gray code. This binary template $\mathbf{t}$ is stored in the system for matching purposes.

EEG is a continuous signal in nature and a moving window of short length is usually applied to segment the data sources into frames for preprocessing and feature extraction. It is natural to use the multiple frames captured during data collection, instead of a single frame, to generate a more stable template. Let $F$ denote the number of frames collected during enrollment or authentication. Each frame corresponds to a vector $\mathbf{r_f}$, hence, $F$ vectors, i.e., $\mathbf{r}_f$ (where $f=1,\cdots, F$), are obtained. The final template generated by the transformation module is the gray encoding of the average of these vectors. The complete transformation procedure is summarized in Algorithm~\ref{alg:transform}. Note that the number of frames used in authentication does not need to be the same as in the registration stage, e.g., 10 frames for generating a template during registration and only 1 for a query sample. The number of frames is adjustable in accordance with application scenarios and requirements. 

\begin{algorithm}[h]
\SetAlgoLined
\caption{Transform}
\label{alg:transform}
\SetKwInOut{Input}{Input}
\SetKwInOut{Output}{Output}
\Input{feature vectors from $F$ frames $\mathbf{v}_{f,1}, \mathbf{v}_{f,2} \in \mathbb{R}^{N+6}, f=1,\cdots,F $ \\
user identity $u$} 
\Output{template $\mathbf{t}$}
\eIf{enrollment}
{initialize a key $k_u$}
{retrieve the key $k_u$}
compute $\mathbf{p} \leftarrow randperm(k_u,N+6)$ \\
compute $\mathbf{M} \leftarrow rand(k_u,[N+6,\delta(N+6)]), \delta \in (0,1)$ \\
\For{$f=1$ \KwTo $F$}
{
permutation $\mathbf{v}'_{f,1} \leftarrow perm(\mathbf{v}_{f,1},\mathbf{p})$ \\
Hadamard product $\mathbf{c}_f \leftarrow \mathbf{v}'_{f,1} \circ \mathbf{v}_{f,2}$ \\
projection $\mathbf{r}_f \leftarrow \mathbf{c}_f \cdot \mathbf{M}$ \\
}
compute $\mathbf{r} \leftarrow (\sum_{f=1}^F \mathbf{r}_f)/F $ \\
encoding $\mathbf{t} \leftarrow GreyCode(\mathbf{r},\text{8-bit})$ 
\end{algorithm}

\subsection{Matching in the Encrypted Domain}

To authenticate a user, one or more frames of EEG signals are captured from the user and a query template is generated following the same procedure as in the registration stage. The Hamming distance is used for calculating the matching score between the query template and the registered template (both are binary representations), as follows:
\begin{equation}
    s(\mathbf{t}_q,\mathbf{t}_r) = sum(\mathbf{t}_q \oplus \mathbf{t}_r),
\end{equation}
where the symbol $\oplus$ denotes element-wise XOR. Finally, the score is compared with a pre-defined threshold $\theta$ to make a decision, as follows:
\begin{equation}
    \hat{o} = 
    \begin{cases} accept, & \text{if }  s(\mathbf{t}_q,\mathbf{t}_r) \leq \theta\\
    reject, & \text{otherwise}
    \end{cases}.
\end{equation}
In the analysis, the threshold $\theta$ is automatically adjusted to obtain the equal error rate (EER), which is defined as the error rate when the false acceptance rate (FAR) equals the false rejection rate (FRR). The FAR reflects the percentage of queries in which impostors are incorrectly accepted, and the FRR reflects the percentage of queries in which genuine users are incorrectly rejected.
\\
\\
\textit{\textbf{Remarks:}} The proposed transformation provides a concise and elegant solution to the generation of secure and cancellable templates. (i) If a template is compromised, the associated user key can be replaced and a new template can therefore be generated with this new key. (ii) Every time the user key is updated, the random permutation in (\ref{eq:permutation}) and the Hadamard product in (\ref{eq:product}) provide a different set of variables for the random projection in (\ref{eq:random_projection}). Since the projection matrix is rank-deficient for every set of variables, it is insufficient to inverse the computation, making the system resistant to the ARM. (iii) The transformation takes advantage of EEG signal elicitation protocol fusion such that the entropy and reliability of the feature vectors are enhanced. (iv) The encoding procedure is a quantization process alleviating the impact of EEG uncertainties associated with the complexity and variability of brain dynamics.

\section{Experimental Design}\label{sec:experiment}

\subsection{Database and Pre-Processing}
The EEG Motor Movement/Imagery Database (MMIDB) from the PhysioNet~\cite{goldberger2000physiobank} is used for evaluating the proposed method. The MMIDB database provides EEG signals collected from 109 subjects in two resting states, EC and EO, and motor imagery tasks including physically opening/closing fists/feet and imagining opening/closing fists/feet without actual body movement. We refer to these two types of motor tasks as physical movement (PHY) and imagery movement (IMA). Since the proposed method adopts the resting state signal elicitation protocol, we use the resting state EEG to evaluate our system. Data acquisition was performed using a BCI2000 system~\cite{schalk2004bci2000} equipped with 64 electrodes with a sampling rate of 160 Hz. The recorded signal is referenced to the earlobes.

For signal pre-processing, an FIR filter with Hamming window was implemented to restrict the frequency range to [0.5 42] Hz after removing the DC offset (the linear trend), then artifacts were removed using independent component analysis and the MARA algorithm~\cite{winkler2014robust}. A moving window of two seconds with no overlapping was used for generating signal frames. Therefore, each frame corresponds to two-second EEG data, with a size of $64\times320$. A template is then generated from each frame by feature extraction and transformation presented in Sections~\ref{sec:feature} and~\ref{sec:transform}.

\subsection{Evaluation Procedures}
To validate the efficacy of each key component of the proposed method, we carry out a comparative study on the signal acquisition protocols, features, transformation, and matching. 

\subsubsection{Signal acquisition protocols}
The dataset provides EEG signals under the resting state and motor imagery tasks. We use the resting state to evaluate the proposed system, and the motor imagery tasks (PHY and IMA) as extra conditions to study the effect of signal acquisition protocols on biometric performance. 

\subsubsection{Feature comparison}
We select three representative features for comparison, which are the reflection coefficients of AR models, band power, and fuzzy entropy. These three types of features capture the time-dependency, power spectral characteristics, and dynamic complexity of EEG signals, respectively. They are classic and important EEG features in time and frequency domains, and have been widely used for EEG biometric applications. In the following analysis, we refer to them as AR, PSD, and FuzzEn, respectively.

\begin{itemize}
    \item AR: An AR model describes the time-varying processes in EEG by specifying that the value of the timeseries at a certain time depends linearly on its own previous values and on a stochastic term (white noise), i.e., $s(t) = \sum_{i=1}^p \theta_i s(t-i) + \varepsilon(t)$, where $\theta$ is the coefficients of the AR model. In this study, we use an AR model of order 5 to fit the signal timeseries, and derive the reflection coefficients as features using the Burg method~\cite{campisi2011brain}. The final feature vector has a length of $5\times N$, where $N$ is the number of channels.  
    \item PSD: We estimate the power spectral density of EEG signals using a non-parametric approach based on the fast Fourier transform. This approach is selected because it directly corresponds to the physical interpretation in terms of EEG rhythms~\cite{la2014human}. Based on the PSD, the average band power over the delta, theta, alpha, beta, and gamma bands are extracted as features. The length of the final feature vector is $5\times N$, where $N$ is the number of channels.
    \item FuzzEn: Entropy quantifies the amount of uncertainty in the EEG amplitudes. Among the existing entropy estimation methods such as approximate entropy and sample entropy, we select the FuzzEn~\cite{mu2016eeg}, which was shown to be a more reliable measure than others for biological data, since the uncertainty at the boundaries between classes can provide a shade of ambiguity~\cite{chen2007characterization}. The final feature vector has a length of $N$, the number of channels. 
\end{itemize}

\subsubsection{Baseline matching performance}
This refers to biometric performance of the system without the transformation module, where template matching is performed in the original feature domain. The non-invertible transformation often needs to reset the order or position of the feature set, which is likely to weaken the discriminant power of the feature set and introduce extra intra-user variations, thus affecting biometric performance~\cite{teoh2008cancellable}. A good cancellable template design should enhance the security of the template without compromising the biometric performance. The baseline performance is used to show whether and to what extent the proposed transformation has an impact on biometric performance.

\subsubsection{Matching performance in the transformed domain}
This refers to biometric performance of the system with the proposed transformation module, where the template is in binary format and matching is conducted in the transformed domain.

\section{Results}\label{sec:result}
There are 109 subjects in the database and each has one minute recording of EC and one minute recording of EO, leading to 30 frames per subject for each state. Depending on how many consecutive frames are used to generate a template during the enrollment and authentication phases, the number of genuine tests varies. Let $F_e$ and $F_t$ denote the number of consecutive frames involved in a template during enrollment and authentication, respectively. We use the first $F_e$ frames for enrollment and the remaining frames for testing. With $1\leq F_e \leq20$ and $F_t=1$, the number of genuine tests varies between 3161 and 1090. One frame from all subjects other than the user is used for the impostor test, resulting in 11772 impostor tests. The authentication performance is measured by the EER.

\subsection{Performance in the Lost Key Scenario}
The lost key scenario is considered the worst case as the user loses his/her parameter key. This means that the attacker can take this advantage to penetrate the authentication system. In order to simulate this scenario, we use the same parameter key to generate the permutation vector $\mathbf{p}$ in (\ref{eq:permutation}) and projection matrix $\mathbf{M}$ in (\ref{eq:random_projection}) for all users in the transformation module. The baseline matching performance of each type of features is summarized in Table~\ref{tbl:eer_baseline}, in which each method was evaluated under four signal elicitation protocols. 

\begin{table}[ht]
\setlength{\tabcolsep}{6pt}
\renewcommand{\arraystretch}{1.3}
\centering
\caption{Baseline matching performance (EER) ($F_e=10$ and $F_t=1$) under different signal elicitation protocols.}
\begin{tabular}{l|llll}
\hline
\multirow{2}{*}{Features} & \multicolumn{4}{c}{Signal elicitation protocols} \\ \cline{2-5}
 & EO & EC & PHY & IMA \\
\hline
AR & 18.27\% & 16.88\% & 20.28\% & 19.38\% \\
PSD & 28.51\% & 30.68\% & 27.27\% & 25.59\% \\
FuzzyEn & 24.54\% & 24.98\% & 22.74\% & 22.42\% \\
AR+PSD+FuzzyEn & 17.40\% & 15.71\% & 15.11\% & 14.80\% \\
Graph (this study) & \textbf{5.19\%} & \textbf{11.27\%} & \textbf{4.39\%} & \textbf{4.93\%} \\
\hline
\end{tabular}
\label{tbl:eer_baseline}
\end{table}

\begin{table*}[ht]
\setlength{\tabcolsep}{10pt}
\renewcommand{\arraystretch}{1.3}
\centering
\caption{Matching performance (EER) in the transformed domain ($F_e=10$ and $F_t=1$) under different signal elicitation protocols.}
\begin{tabular}{l|cccc|cc}
\hline
\multirow{2}{*}{Features} & \multicolumn{4}{c|}{Signal elicitation protocols} & \multicolumn{2}{c}{Fusion embedded in the transformation module} \\ \cline{2-7}
 & EO & EC & PHY & IMA & EO+EC & PHY+IMA \\
\hline
AR & 22.76\% & 23.99\% & 24.17\% & 22.64\% & 18.52\% & 18.43\% \\
PSD & 28.69\% & 26.66\% & 23.43\% & 22.55\% & 21.50\% & 13.78\% \\
FuzzyEn & 31.77\% & 32.86\% & 32.91\% & 32.27\% & 25.65\% & 29.22\% \\
AR+PSD+FuzzyEn & 18.05\% & 20.60\% & 15.97\% & 14.54\% & 14.46\% & 11.53\% \\
Graph (this study) & 11.31\% & 15.90\% & 8.29\% & 8.71\% & \textbf{8.58\%} & \textbf{5.98\%} \\
\hline
\end{tabular}
\label{tbl:eer_transform}
\end{table*}

\begin{figure}[h]
\centering
\includegraphics[width=0.33\textwidth]{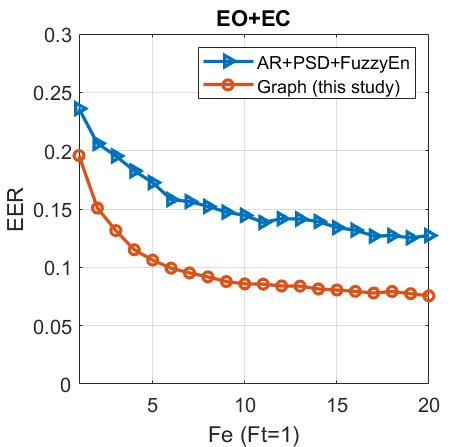}
\caption{Matching performance (EER) of the system in the transformed domain using a varying number of frames for template enrollment ($F_e \in \{1,2,\cdots,20\}$ and $F_t=1$) with resting state protocol fusion.}
\label{fig:eer_transform}
\end{figure}

Table~\ref{tbl:eer_transform} summarizes the matching performance of each type of features in the transformed domain, in which each method was evaluated under four signal elicitation protocols and two fusion conditions. In addition, Fig.~\ref{fig:eer_transform} shows the performance of the system using a varying number of frames for the enrolled template, under the proposed transform-embedded protocol fusion. From Tables~\ref{tbl:eer_transform} and~\ref{tbl:eer_baseline}, we can observe that the proposed cancellable template design (graph features + transformation) demonstrates comparable authentication performance while protecting original EEG biometrics. 

\begin{figure*}[h]
\centering
\includegraphics[width=0.98\textwidth]{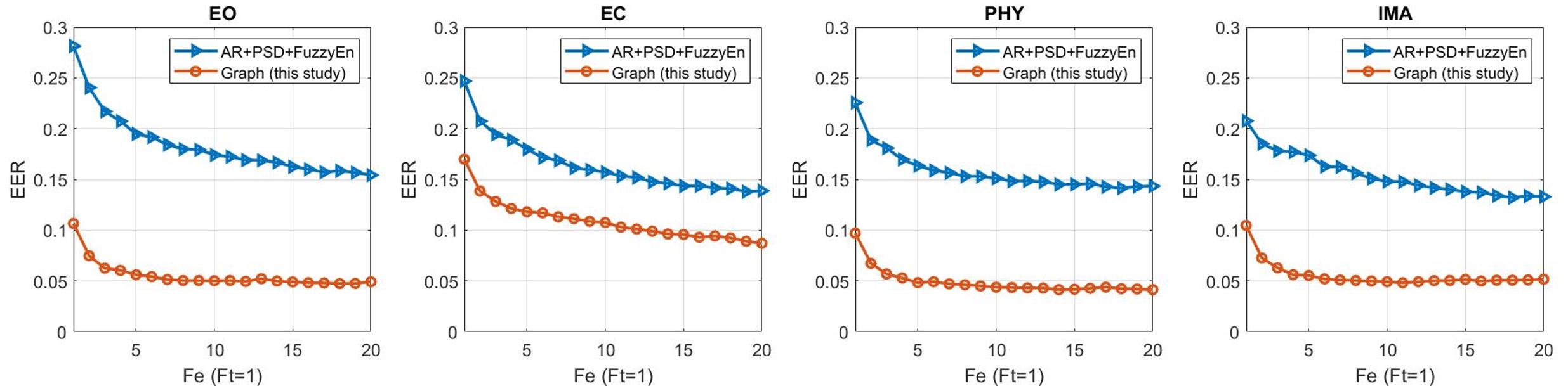}
\caption{Baseline matching performance (EER) of the system using a varying number of frames during template enrollment ($F_e \in \{1,2,\cdots,20\}$ and $F_t=1$) under different signal elicitation protocols.}
\label{fig:eer_baseline}
\end{figure*}

In Table~\ref{tbl:eer_transform}, comparing the results of elicitation fusion (embedded in the transformation) with those of single elicitation protocols, we can see a clear improvement in the authentication performance, which shows the effectiveness of embedding the protocol fusion in the transformation for enhancing the authentication performance. The results also show that, although the proposed method uses resting state protocols and graph features, the transformation itself is not confined to specific signal elicitation protocols or features. Furthermore, we evaluate the effect of the number of frames contained in the registration template on the biometric performance (EER). The results are reported in Fig.~\ref{fig:eer_baseline} and Fig.~\ref{fig:eer_transform}. A lower EER is achieved when increasing the number of frames for template generation. The trend is consistent in both the transformed and non-transformed domains. This is reasonable because having more EEG frames helps enhance the reliability of the template. 

\subsection{Decidability Analysis}
\begin{figure}[h]
\centering
\includegraphics[width=0.98\columnwidth]{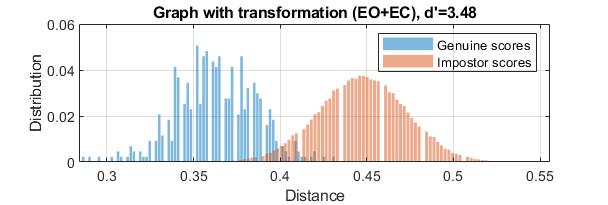}
\includegraphics[width=0.98\columnwidth]{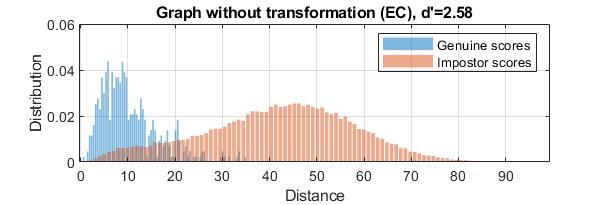}
\includegraphics[width=0.98\columnwidth]{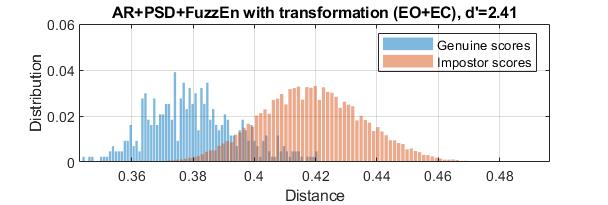}
\includegraphics[width=0.98\columnwidth]{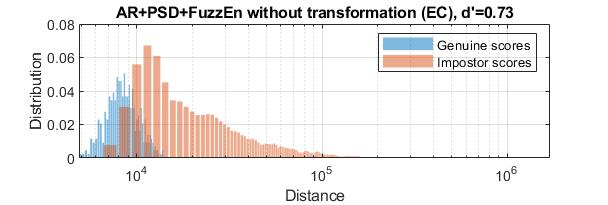}
\caption{Decidability analysis results. Distributions of the genuine scores and impostor scores, and the corresponding decidability index $d'$, demonstrated for User 5.}
\label{fig:score_distribution}
\end{figure}

Biometric authentication is essentially a decision task to discriminate the user from impostors. In this analysis, we adopt the decidability index $d'$~\cite{williams1996use} to measure the discriminant capacity of the designed cancellable template. The $d'$ is defined as:
\begin{equation}
    d'= (m_{intra} - m_{inter}) / \sqrt{(s_{intra}^2+s_{inter}^2)/2}
\end{equation}
where ($m_{intra}$, $s_{intra}$) and ($m_{inter}$, $s_{inter}$) denote the mean and standard deviation of the matching scores between user samples and the matching scores between user samples and impostor samples, respectively. For each user, we generate a genuine score distribution by matching every possible pair of the user samples, and an impostor score distribution by matching each user sample with each sample of other subjects, which yields 435 genuine scores and 97200 impostor scores. Fig.~\ref{fig:score_distribution} presents the score distributions of the proposed method, i.e., Graph+transform(EO+EC), and three comparison methods, namely Graph features without the transformation, and AR+PSD+FuzzyEn features with and without the transformation. The observation is that the proposed transformation enhances the decidability of the system: from 2.58 to 3.48 for graph features and from 0.73 to 2.41 for classical features (AR+PSD+FuzzyEn). As can be noticed from the figure, the proposed transformation reduces the overlap between the genuine and impostor score distributions.

\subsection{Revocability and Diversity}
\begin{figure}[h]
\centering
\includegraphics[width=0.98\columnwidth]{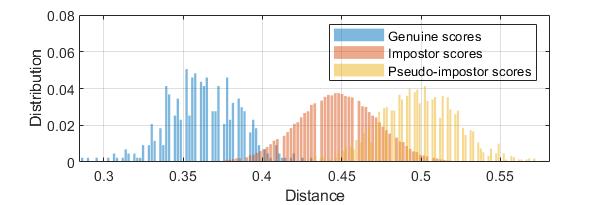}
\caption{Revocability and diversity analysis results. Distributions of the pseudo-impostor score, genuine scores, and impostor scores.}
\label{fig:pseudo_impostor}
\end{figure}

The revocability and diversity criteria specify that templates generated from the same biometric features by different parameter keys should have no correlation. To evaluate the capacity of the proposed cancellable template design in terms of revocability and diversity, we follow the common practice in relevant studies~\cite{tran2021multi} and calculate the pseudo-impostor scores. For each user, 50 additional transformed templates (i.e., the pseudo-impostor) are generated from the first feature template using different parameter keys. A pseudo-impostor score distribution can then be obtained by matching the original user templates with the pseudo-impostor templates of the same user. Fig.\ref{fig:pseudo_impostor} shows the pseudo-impostor score distribution of the proposed method, along with the genuine and impostor score distributions. The corresponding statistics of these scores are reported in Table~\ref{tbl:scores}. The results show that the pseudo-impostor score distribution has almost no overlap with the genuine score distribution. The statistics also show that there is a clear difference between genuine and impostor scores, and between genuine and pseudo-impostor scores. Meanwhile, the pseudo-impostor score distribution is significantly overlapped with the impostor score distribution. In other words, the system satisfies the revocability and diversity requirements.

\begin{table}[ht]
\setlength{\tabcolsep}{10pt}
\renewcommand{\arraystretch}{1.3}
\centering
\caption{Mean and standard deviation of the pseudo-impostor score, genuine scores, and impostor scores.}
\begin{tabular}{l|c|c}
\hline
Scores & Mean & Std \\
\hline
Genuine user & 0.363 & 0.023 \\
Impostor & 0.447 & 0.025 \\
Pseudo-impostor & 0.499 & 0.025 \\
\hline
\end{tabular}
\label{tbl:scores}
\end{table}

\subsection{Unlinkability}
\begin{figure}[h]
\centering
\includegraphics[width=0.7\columnwidth]{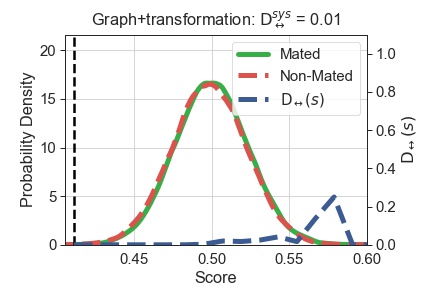}
\includegraphics[width=0.7\columnwidth]{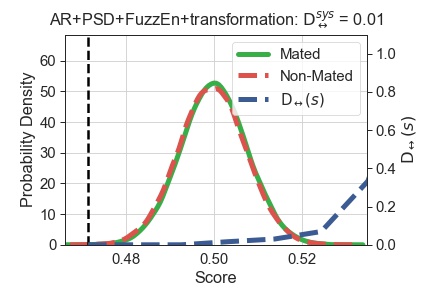}
\caption{Unlinkability analysis results. The mated score and non-mated score distributions and the two linkability measures: local $D_{\leftrightarrow}$(s) and global $D_{\leftrightarrow}^{sys}$.}
\label{fig:linkability}
\end{figure}

For a cancellable biometric template design, the unlinkability property requires that the transformed templates originated from the same EEG data of the same subject are as different as those from different subjects~\cite{gomez2017general}. To evaluate the unlinkability of the proposed method, we adopted two measures, i.e., the score-wise linkability $D_{\leftrightarrow}(s)$ (a local measure) and system overall linkability $D_{\leftrightarrow}^{sys}$ (a global measure)~\cite{gomez2017general}, which are popular tools for unlinkability assessment in cancellable biometrics research. The calculation of $D_{\leftrightarrow}(s)$ and $D_{\leftrightarrow}^{sys}$ is based on the mated and non-mated sample score distributions. The mated sample score is obtained by matching two templates generated from the same EEG data using different parameter keys. The non-mated sample score is obtained by matching two templates generated from the EEG of different subjects using different parameter keys. We followed the same procedure in a recent study~\cite{tran2021multi} and generated six transformed databases using six different keys. The value range of $D_{\leftrightarrow}(s)$ and $D_{\leftrightarrow}^{sys}$ is [0, 1] with 0 indicating fully unlinkable and 1 indicating fully linkable. Fig.~\ref{fig:linkability} presents the analysis results, where we tested two methods, the proposed method `Graph + transformation', and the proposed transformation with classical features `Classic features + transformation'. It is clear from Fig.~\ref{fig:linkability} that the proposed method provides high unlinkability, with a very low global linkability index $D_{\leftrightarrow}^{sys}=0.01$. It also indicates that the proposed transformation is effective and able to work with different features, e.g., classic features (AR+PSD+FuzzyEn).

\section{Security Analysis and Discussion}~\label{sec:security}

\subsection{Attacks via Record Multiplicity (ARM)}
A cancellable biometric template design allows distinct transformed templates $\{\mathbf{y}_1,\mathbf{y}_2,\cdots,\mathbf{y}_n\}$ to be generated from the same raw biometric template $\mathbf{x}$ by applying different transformation parameters $\{k_1,k_2,\cdots,k_n\}$. ARM refers to the attack that aims to retrieve the raw biometric template $\mathbf{x}$ by correlating multiple transformed templates $\{\mathbf{y}_1,\mathbf{y}_2,\cdots,\mathbf{y}_n\}$, assuming that these transformed templates as well as information about the transformation method $F$ and corresponding parameters $\{k_1,k_2,\cdots,k_n\}$ are available~\cite{li2014attacks}. 

The resistance of the proposed method to the ARM attack is guaranteed by the non-invertible transformation with three key points, i.e., random projection, random permutation and Hadamard product. First of all, the random projection procedure $\mathbf{y} = \mathbf{x} \cdot \mathbf{M}$ in (\ref{eq:random_projection}) provides one-time-pad security so that each individual $\mathbf{y}_i$ cannot be reversed to obtain $\mathbf{x}$, as proved and demonstrated in previous studies~\cite{wang2012alignment}. However, the random projection itself is exposed to ARM because a unique solution can be determined by solving a well-defined system of linear equations $\{\mathbf{y}_i = F(\mathbf{x},k_i)\}, i=1,2,\cdots,n$. To address this issue, the random permutation and Hadamard product operation is performed before the random projection. Note that the input of the random projection is actually the Hadamard product of $\mathbf{v}_2$ and permutation of $\mathbf{v}_1$ as in $\mathbf{c} = perm(\mathbf{v}_1,\mathbf{p}) \circ \mathbf{v}_2$, where $\mathbf{v}_1$ and $\mathbf{v}_2$ are two feature vectors; see (\ref{eq:permutation}) and (\ref{eq:product}). For different values of $k$ in (\ref{eq:permutation}), the random permutation and Hadamard product would produce different sets of variables for the random projection-based linear equations, thus a well-defined system of linear equations cannot be established. Since the projection matrix $\mathbf{M}$ is rank-deficient for every set of variables, it is insufficient to inverse the computation in (\ref{eq:random_projection}). 

Below is a representative example to show how the proposed method protects the system from the ARM attack. For demonstration purposes, we use low dimensional real vectors $\mathbf{v}_1=[v_{11}, v_{12}, v_{13}, v_{14}]$ and $\mathbf{v}_2=[v_{21}, v_{22}, v_{23}, v_{24}]$ to represent the real-valued feature vectors under protection. Suppose that the feature vectors are $\mathbf{v}_1 = [0.19, 0.54, 0.37, 0.84]$ and $\mathbf{v}_2 = [0.59, 0.18, 0.04, 0.92]$. Given two transformation parameters $k_1 = 1$ and $k_2 = 10$, we can produce two sets of permutations $\mathbf{p}_1=[3, 4, 1, 2]$ and $\mathbf{p}_2=[2, 3, 4, 1]$ and projection matrices $\mathbf{M}_1=[0.15, 0.40; 0.09, 0.54; 0.19, 0.42; 0.35, 0.69]$ and $\mathbf{M}_2=[0.50, 0.17; 0.22, 0.09; 0.20, 0.69; 0.76, 0.95]$. Applying the proposed transformation, we can get two transformed templates $\mathbf{t}_1=[0010001110100010]$ and $\mathbf{t}_2=[1010001000100011]$, which are the codes of real vectors $ \mathbf{r}_1=[0.22, 0.51]$ and $\mathbf{r}_2=[0.31, 0.25]$, respectively. Now, suppose that an adversary gets $\mathbf{t}_1$, $\mathbf{t}_2$, $k_1$, $k_2$, knows the transformation function, and wants to retrieve $\mathbf{v}_1$ and $\mathbf{v}_2$. The first step taken by the adversary would be to decode the binary templates into the corresponding real values, which can be possible in the worse case, assuming that the adversary is able to collect massive amounts of encoded data and get the distribution of the values through statistical tools. Suppose that the estimated real vectors are $\hat{\mathbf{r}}_1 = [0.2, 0.5]$ and $\hat{\mathbf{r}}_2 = [0.3, 0.2]$, then the key step is to solve the following equations:
\begin{equation*}
\begin{cases}
0.2&= 0.15 v_{13}v_{21} + 0.09 v_{14}v_{22} + 0.19 v_{11}v_{23} + 0.35 v_{12}v_{24}\\
0.5&= 0.4 v_{13}v_{21} + 0.54 v_{14}v_{22} + 0.42 v_{11}v_{23} + 0.69 v_{12}v_{24}\\
0.3&= 0.5 v_{12}v_{21} + 0.22 v_{13}v_{22} + 0.2 v_{14}v_{23} + 0.76 v_{11}v_{24}\\
0.2&= 0.17 v_{12}v_{21} + 0.09 v_{13}v_{22} + 0.69 v_{14}v_{23} + 0.95 v_{11}v_{24}
\end{cases}
\end{equation*}

However, the above is an ill-posed problem in that there is no unique solution and the solution is highly sensitive to changes in the estimated $\hat{\mathbf{r}}$. Using Matlab pseudo-inverse function, we get $\hat{\mathbf{v}}_1 = [0.39, 0, 0.85, 0.16]$ and $\hat{\mathbf{v}}_2 = [1, 0, 1, 0]$. However, the cosine similarity of the estimated value and ground truth value is $\frac{\mathbf{v}\hat{\mathbf{v}}^\intercal}{ {\|\mathbf{v}\|}_2 {\|\hat{\mathbf{v}}\|}_2} = 0.44$, which indicates that the obtained value is far from the true biometric data. Our analysis shows that if a transformed template stored in the database is compromised, it reveals no clue about the original biometric data. Even in the worst-case scenario where multiple sets of templates and the corresponding parameter keys are exposed, it would be highly unlikely to retrieve the true biometric data from infinite solutions. 
 
\subsection{Pre-image Attacks}
The original definition of a pre-image attack on a cryptographic hash function refers to an attacker trying to determine an input that has a specific hash value. A cryptographic hash function $f(\cdot)$ should resist attacks on its pre-image, that is, given $y$, it is difficult to find $x$ such that $y = f(x)$. Such a definition does not fully apply in the context of transformation-based (i.e. non-cryptographic) cancellable design schemes. Strictly speaking, for a transformation-based cancellable design, it is possible to find an input $x$ given $y$ such that $y = f(x)$. However, it will be of little value if the solution $x$ is not the original biometric feature under protection and the compromised template is revoked. Considering the properties of transformation-based cancellable schemes, we therefore redefine the pre-image attack as follows:

\textit{Given a transformed template $\mathbf{y}$, it is difficult to find a solution $\mathbf{x}$ such that $\mathbf{y}=f(\mathbf{x},K)=f(\mathbf{x}_0,K)$ and $\mathbf{x}=\mathbf{x}_0$, where $f(\cdot)$ is the transformation function with parameter key $K$, and $\mathbf{x}_0$ is the original biometric feature.}

The proposed transformation is a many-to-one mapping function, and we have demonstrated in the ARM attack analysis that it would be difficult to find the real input in a systematical way. In the following hill-climbing attack analysis, Case I can also be considered as a pre-image attack. We will show that the solution found by the hill-climbing attack is far away from the real input, and therefore, the solution becomes insignificant once the compromised template is revoked.

\subsection{Hill-Climbing Attacks}
\begin{figure}[h]
\centering
\includegraphics[width=0.98\columnwidth]{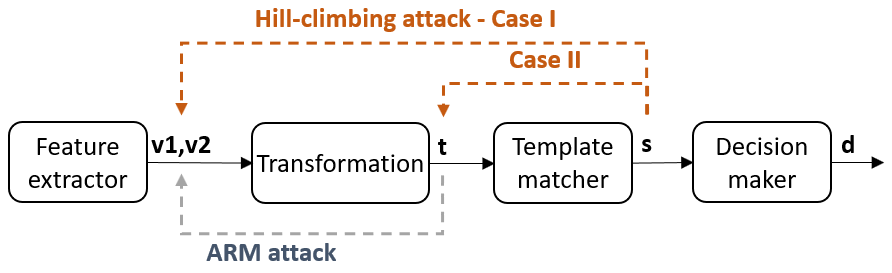}
\caption{Hill-climbing attacks on the system.}
\label{fig:hill-climbing}
\end{figure}

This refers to an adversary exploiting the matching scores to generate synthetic biometric data that would allow a false acceptance~\cite{maiorana2013vulnerability}. In the context of cancellable biometrics, the hill-climbing attack can be launched in two ways, as illustrated in Fig.~\ref{fig:hill-climbing}. Case I -- the adversary submits and tries to obtain feature vectors $\mathbf{v}_1$ and $\mathbf{v}_2$ as in the conventional non-cancellable context~\cite{maiorana2013vulnerability}. Case II -- the adversary submits and tries to obtain template $\mathbf{t}$ stored in the system.  Hill-climbing attacks are a threat to conventional non-cancellable biometric systems as the adversary is able to get a synthetic feature vector that is very close to the true feature vector and compromise the system with it. However, this is not necessarily true for cancellable biometric systems. In the following, we will demonstrate that cancellable biometric systems, especially those based on many-to-one mapping, are naturally resistant to hill-climbing attacks. 

\begin{figure*}[h]
\centering
\includegraphics[width=0.3\linewidth]{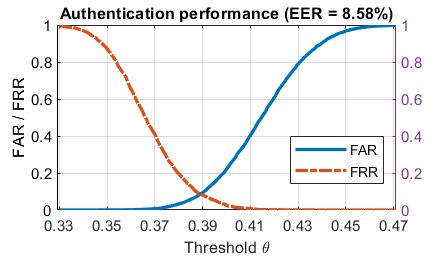}
\includegraphics[width=0.3\linewidth]{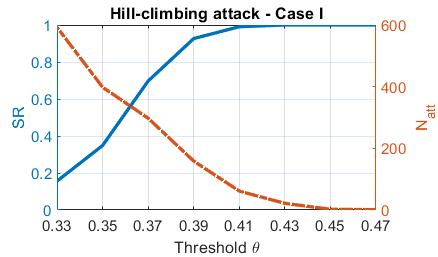}
\includegraphics[width=0.3\linewidth]{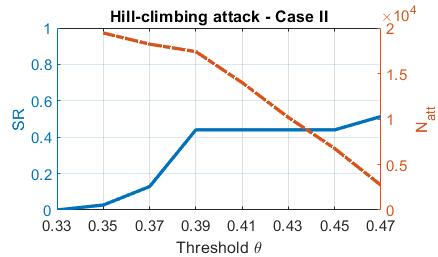}
\caption{The SR and $N_{att}$ of the hill-climbing attack on the system.}
\label{fig:hill-climbing_res}
\end{figure*}

The Nelder-Mead algorithm was used to implement the hill-climbing attack. It is a downhill simplex method that is among the most well-known algorithms for derivative-free optimization~\cite{maiorana2013vulnerability}. The evaluation of the objective function $\mathcal{F(\cdot)}$ represents the difference between the input query and the reference template. The process ends either when the minimum value of the objective function is equal to or less than the system threshold (here we set $\theta=0.389$, i.e., the EER operating point) or when the maximum number of attempts is reached (here set to 20,000). The system's vulnerability to hill-climbing attacks is measured by the success rate (SR), defined as the percentage of users whose accounts are compromised within 20,000 attempts. The efficiency of the attack is measured by $N_{att}$, the average number of attempts required to successfully crack an account. Results of the hill-climbing attack are presented in Fig.~\ref{fig:hill-climbing_res}. We can see that it is possible to find a solution to temporarily break in user accounts with hill-climbing attacks. At the EER operating point, the SR of hill-climbing attacks is around 0.899 and 0.358 in Cases I and II, respectively. It is also worth noting that when adjusting the system operating threshold towards a lower FAR, the SR and efficiency of launching hill-climbing attacks decrease significantly. 

Now, let us assume the adversary has successfully found a solution to pass the system through the hill-climbing attack. We will demonstrate that this solution will fail once the system changes the cancellable template. Let $\mathbf{t}_0$ denote the transformed template stored in the system before attack; and $\hat{\mathbf{v}}_1$, $\hat{\mathbf{v}}_2$, and $\hat{\mathbf{t}}_0$ denote the feature vectors and template obtained through the hill-climbing attack.

In Case I, the adversary obtained an estimated solution $\hat{\mathbf{v}}_1$ and $\hat{\mathbf{v}}_2$ that generates a template close enough to $\mathbf{t}_0$ to pass the system. To defend, the system will replace the compromised template $\mathbf{t}_0$ with a new one $\mathbf{t}_1$ using a new set of transformation parameters. Let $\hat{\mathbf{t}}_1$ denote the query template generated from the estimated $\hat{\mathbf{v}}_1$ and $\hat{\mathbf{v}}_2$ with the same new transformation parameters. We now demonstrate that $\hat{\mathbf{t}}_1$ is not a valid solution for $\mathbf{t}_1$. In our experiment, there were 98/109 users (SR=0.899 at operating point of EER) whose original templates were successfully attacked by hill-climbing attacks. For each of those 98 users, we replaced the compromised template with a new one using a new key and tested whether the query generated from the obtained solution using the new key is able to match the new template. To have a reliable analysis, we randomly generated 200 keys for each user, which yields a total number of 19600 (98$\times$200) tests. The results in Table~\ref{tbl:hill-climbing} show that it is highly unlikely (0/19600) that the $\hat{\mathbf{t}}_1$ (generated from the adversary's obtained solution) can match the corresponding true template $\mathbf{t}_1$ (generated from the true feature vectors) to break in the account. A further examination shows that the hill-climbing solution $\{\hat{\mathbf{v}}_1, \hat{\mathbf{v}}_2\}$ is far away from the true feature vectors $\{\mathbf{v}_1, \mathbf{v}_2\}$, with a very low similarity score of 0.193 $\pm$ 0.004. This is because the proposed transformation is a many-to-one mapping and irreversible function, thus it is hard to hit the true solution through hill-climbing. Our results prove that even if a temporary solution is found, it is unlikely that this fake solution can pass the system after the compromised template is revoked.

In Case II, the adversary obtained $\hat{\mathbf{t}}_0$, which is an approximation of $\mathbf{t}_0$. To defend, the system can simply replace the compromised $\mathbf{t}_0$ with a new template $\mathbf{t}_1$ using a new set of transformation parameters. In our experiment, there were 39/109 users (SR=0.358 at the operating point of EER) whose original templates were successfully attacked by hill-climbing attacks. For each of those 39 users, we replaced the compromised template with a new one using a new key and tested whether the obtained solution can match the new template to break in the account again. To have a reliable analysis, we randomly generated 200 keys for each user, which yields a total number of 7800 (39$\times$200) tests. The results in Table~\ref{tbl:hill-climbing} show that the hill-climbing solution $\hat{\mathbf{t}}_0$ is not similar to any of these new templates (with a low similarity score of 0.219 $\pm$ 0.036), and none of them can successfully match the new templates. Since $\hat{\mathbf{t}}_0$ neither reveals clues about the raw biometric data nor correlates with the new template, obtaining $\hat{\mathbf{t}}_0$ through hill-climbing attacks would be meaningless.

\begin{table}[ht]
\setlength{\tabcolsep}{2pt}
\renewcommand{\arraystretch}{1.3}
\centering
\caption{Results from attacking the system using hill-climbing solutions after the compromised templates are revoked.}
\begin{tabular}{l|c|c|c}
\hline
 & \multirow{2}{*}{Solution similarity}  & \multirow{2}{*}{Matching scores (distance)}  &\#Success/\#Tests \\
 & & & (SAR) \\
\hline
Case I & 0.193 $\pm$ 0.004 & 0.489 $\pm$ 0.025 & 0 / 19600 \\
Case II & 0.219 $\pm$ 0.036 & 0.781 $\pm$ 0.036 & 0 / 7800\\
\hline
\end{tabular}
\label{tbl:hill-climbing}
\end{table}

\subsection{Second Attack Rate - an Extension to the Classical Lost Key Scenario Analysis}
Spawned from hill-climbing attacks is a new concept, which we call \textit{Second Attacks}. It is defined as an attempt to break into a system using pre-obtained solutions after the system has revoked the compromised templates. Accordingly, the \textit{Second Attack Rate (SAR)} is the success rate of these second attack attempts. Here we examine three ways to obtain a valid solution to break in a user account.
\begin{itemize}
\item Mathematical solutions. Assume that the attacker acquires the user template $t$ and knows the transformation function $F$, then it is possible to find a solution $\hat{x}$ such that $t=F(x)$.
\item Computational solutions via hill-climbing attacks. Assume that an attacker can submit $x$ to the system and get the corresponding matching score, then it is possible to find $\hat{x}$ that can pass the system.
\item Public data solutions. Assume that the attacker has a public database $X$, then for a system with a non-zero FAR, it is likely to find $\hat{x}$ that can pass the system by testing each data.
\end{itemize}

Table~\ref{tbl:second_attack} summarizes the performance of the proposed cancellable biometrics design in terms of SAR using the aforementioned three types of solutions. An interesting finding is that the computational solution obtained through the hill-climbing attack is not better than the public data solution or the mathematical solution. This finding validates our hypothesis from an experimental point of view that a good cancellable biometrics design based on many-to-one mapping is inherently resistant to hill-climbing attacks.

\begin{table}[ht]
\setlength{\tabcolsep}{2pt}
\renewcommand{\arraystretch}{1.3}
\centering
\caption{SAR results under three types of solutions.}
\begin{tabular}{l|c|c|c|c}
\hline
 Solution & Solution similarity & Matching score (dist.) & \#Tests  & SAR\\
\hline
Mathematical & 0.167 $\pm$ 0.129 & 0.497 $\pm$ 0.025 & 109$\times$200 & 0\\
Computational & 0.193 $\pm$ 0.004 & 0.489 $\pm$ 0.025 & 98$\times$200 & 0\\
Public data & 0.033 $\pm$ 0.005 & 0.499 $\pm$ 0.025 & 109$\times$200 & 0\\
\hline
\end{tabular}
\label{tbl:second_attack}
\end{table}

\subsection{Entropy and Brute Force Attacks}
A brute force attack attempts to guess the elements of the original EEG feature vectors $\mathbf{u}$ and $\mathbf{v}$ through exhaustive search. We analyze the search space to show the likelihood of finding the secret successfully from the search space. In the proposed system, both $\mathbf{u}$ and $\mathbf{v}$ have a length of $m\times n$ bits, where we have $m=70$ and $n=8$ in our experimental setup. Therefore, the number of trials needed to attack the system would be $2^{1120}$, which is computationally expensive. In actual deployment, the settings of $m=70$ and $n=8$ can be adjusted according to the requirements. For example, a larger $m$ can further enhance the security level, however, it is worth noting that a larger $m$ also implies a higher computational cost or less efficiency in data collection. There is a trade-off between performance, security and system efficiency.

\subsection{Pitfalls in the evaluation procedure of supervised learning-based authentication systems}
EEG biometrics has become a hot topic in recent years, and research articles in this field have sprung up in large numbers. Among these papers, many treat authentication as a pure classification problem without considering the differences between the two concepts. In the following analysis, we take the LDA and SVM, two popular classifiers widely used in EEG biometrics, as an example to show that user authentication is not merely a classification problem, and the standard evaluation procedure for classification systems based on supervised learning (SL) models are not fully applicable to the evaluation of user authentication systems. Assuming that the database has $N$ subjects, the standard evaluation procedure for SL-based systems is illustrated in Algorithm~\ref{alg:standard}.
\begin{algorithm}[h]
\small
\SetAlgoLined
\caption{Standard evaluation for classification}
\label{alg:standard}
predictions = [] \\
\For{$n=1$ \KwTo $N$}
{
 re-label data of subject $n$ as 1 (user) \\
 re-label data of other subjects as 0 (intruder) \\
 train-test-split (80\%, 20\%) (or 5-fold cross-validation) \\
 SL model $\leftarrow$ SL model training on train set \\
 predictions $\leftarrow$ SL model testing on test set \\
}
accuracy $\leftarrow$ confusion matrix (predictions, ground truth)
\end{algorithm}

In the context of user authentication, the evaluation process described above has two major issues. 1) Training a binary SL model requires samples from both classes, and the way it splits train/test sets has provided the model with all intruders' data during the training phase. This is incorrect because no data from any test intruder should be seen by the authentication system until the system is tested. The correct procedure should separate the intruder set from the user set, as applied in~\cite{riera2008unobtrusive}. Adopting the basic idea of separating user and intruder sets, we suggest a more reliable evaluation procedure for EEG-based authentication system (i.e., Algorithm~\ref{alg:evaluation}). 2) Authentication systems based on SL classification models (e.g., LDA and SVM) do not have a system operating threshold. An individual classification model is trained for each user, and the overall system performance is embodied as a pair of FAR and FRR. The ROC curve and EER are reported in some studies, but such ROC or EER is a measure to examine the output of the classifier, not the ROC or EER of the authentication system.
\begin{algorithm}[h]
\small
\SetAlgoLined
\caption{Evaluation procedure for supervised learning-based authentication}
\label{alg:evaluation}
Split database into user set $U$ and intruder set $I$ \\
IntruderTests $\leftarrow$ intruder set $I$ \\
predictions = [] \\
\For{$u=1$ \KwTo $U$}
{
 re-label data of subject $u$ as 1 (user) \\
 re-label data of other subjects as 0 (non-user) \\
 train-test-split (80\%, 20\%) (or 5-fold cross-validation) \\
 SL model $\leftarrow$ SL model training on train set\\
 UserTests $\leftarrow$ testing data of label 1  \\
 predictions $\leftarrow$ SL model testing on UserTests \\
 predictions $\leftarrow$ SL model testing on IntruderTests \\
}
accuracy $\leftarrow$ confusion matrix (predictions, ground truth)
\end{algorithm}

Table~\ref{tbl:classification} reports the results of PSD features with LDA and SVM classifiers under the two evaluation procedures in different settings. The first observation is that the standard classification evaluation procedure gives false high performance, especially the FAR, because it erroneously feeds intruder data to the model during the training phase. In addition, the performance of classification-based authentication systems relies on a good training set. With less training data (a smaller number of users in the system dropped from 80 to 30), the performance would degrade. Another concern is that the 80\%-20\% data split, a very common setup in classification tasks, may be too lax for testing an authentication system. For example, given a database where each subject has 60 samples, the 80-20 split means 48 samples are used for registration (which would take a while for data collection) and only 12 samples are used for positive testing. Our results show that reducing the split ratio (from 80\% to 33\%) also degrades performance. We hope to use this demonstration to rectify the misconception of many existing studies evaluating classifier-based authentication systems. 

\begin{table}[ht]
\setlength{\tabcolsep}{2pt}
\renewcommand{\arraystretch}{1.3}
\centering
\caption{Results (\%) under different evaluation procedures for classification-based systems.}
\begin{tabular}{llllll}
\hline
Method & Evaluation & Accuracy & FAR & FRR & EER (classi.) \\
\hline
PSD+SVM & Classi. (80\%) & 99.56 & 0.37 & 7.80 & 1.68 \\
PSD+SVM & Authen. (80\%, 80 users) & 96.73 & 3.23 & 8.75 & 4.06 \\
PSD+SVM & Authen. (33\%, 80 users) & 96.34 & 3.44 & 12.94 & 4.7 \\
PSD+SVM & Authen. (33\%, 30 users) & 95.37 & 4.58 & 10.33 & 5.89 \\
\hline
PSD+LDA & Classi. (80\%) & 97.16 & 2.76 & 11.16 & 7.80 \\
PSD+LDA & Authen. (80\%, 80 users) & 91.46 & 8.52 & 11.46 & 9.58 \\
PSD+LDA & Authen. (33\%, 80 users) & 89.74 & 10.09 & 17.38 & 12.91 \\
PSD+LDA & Authen. (33\%, 30 users) & 77.51 & 22.52 & 18.50 & 21.65 \\
\hline
\multicolumn{6}{l}{The number in parentheses indicates the split ratio for training set.}
\end{tabular}
\label{tbl:classification}
\end{table}

\section{Conclusion}\label{sec:conclusion}
In this study, we proposed a cancellable biometrics scheme for privacy-preserving EEG-based authentication systems. To be specific, an innovative non-invertible transformation was designed to generate cancellable templates from EEG graph features while taking advantage of signal elicitation protocol fusion to enhance biometric performance. The results demonstrated that the proposed method provides good authentication performance (8.58\% EER), prevents the leakage of the sensitive information contained in the EEG data, and is secure against the ARM attack, pre-image attack, hill-climbing attack, and brute force attack. In particular, we examined two ways to perform hill-climbing attacks, and demonstrated that the solution found through hill-climbing attacks would fail once the system revokes the compromised template. In other words, cancellable biometric systems, especially those based on many-to-one mapping, are naturally resilient against hill-climbing attacks. We also introduced the concept of second attacks for cancellable biometric systems. Finally, we discussed the evaluation procedure of supervised learning-based authentication systems and the pitfalls involved. Our future work will further this line of research to explore the possibility of integrating cryptographic schemes into authentication systems.

\bibliographystyle{IEEEtran}
\bibliography{ref}

\end{document}